\documentclass[12pt,a4paper]{article}
\usepackage{color}
\usepackage{t1enc}
\usepackage[dvips,final]{graphicx} %<--ked chces vkladat obrazky
\usepackage{tabularx} %<--ked chces robit nieco s tabulkami
\usepackage{amssymb,amsfonts,amsmath,euscript}
\usepackage{graphics}
\usepackage{psfrag}
\usepackage{caption}
\usepackage{amsthm}

\newcommand\curl{\text{\rm curl\,}}
\newcommand\Deg{{\text{\rm deg\,}}}
\newcommand\Div{{\text{\rm div\,}}}

\newcommand\sgn{\text{\rm sgn\,}}

\begin{document}
\renewcommand{\figurename}{\footnotesize{Obr.}}
\title{Galilean and Carrollian Hodge star operators}
\author{Mari\'an Fecko\thanks{e-mail: {\texttt{Marian.Fecko@fmph.uniba.sk}}} \\
\small{{Department of Theoretical Physics,}}                                \\
\small{{Faculty of Mathematics, Physics and Informatics,}}                  \\
\small{{Comenius University in Bratislava,}}                                \\
\small{{Mlynsk\'a dolina F2, 84248 Bratislava, Slovakia}}}
%\small{{Marian.Fecko@fmph.uniba.sk}}}
%\author{Mari\'an Fecko\thanks{e-mail: {\texttt{fecko@fmph.uniba.sk}}} \\
%\small{{Department of Theoretical Physics, Comenius University in Bratislava}}}
\date{}
\maketitle
\small{\emph{The standard Hodge star operator is naturally associated with metric tensor (and orientation).
             It is routinely used to concisely write down physics equations on, say, Lorentzian spacetimes.
             On Galilean (Carrollian) spacetimes, there is no canonical (nonsingular) met\-ric tensor available.
             So, usual construction of the Hodge star does not work.
             Here we propose analogs of the Hodge star operator on Galilean (Carrollian) spacetimes.
             They may be used to write down important physics equations,
             e.g. equations of Galilean (Carrollian) electrodynamics.
             }

%\pacs{02.40.-k, 45.20.Jj, 47.10.Df, 47.32.C-}% PACS, the Physics and Astronomy
                             % Classification Scheme.
                             %diff.geometry, Hamiltonian mechanics, vortex dynamics (fluid flow)
%\keywords{Galilean spacetime, Carrollian spacetime, Galilean electrodynamics, Carrollian electrodynamics, Hodge star operator}
                              %Use showkeys class option if keyword
                              %display desired

\vglue -1 cm
\tableofcontents

%\newpage
\vskip 2cm

\section{Introduction - standard Hodge star operator}
\label{intro}

The \emph{Hodge star} operator $*\equiv *_{g,o}$ is a well-known \emph{canonical} linear mapping
acting on differential forms on (pseudo)Riemannian oriented manifold $(M,g,o)$,
\begin{equation} \label{defHodge1}
                 *:\Omega^p \to \Omega^{d-p},
                 \hskip 1cm
                 \Omega^p \equiv \Omega^p(M,g,o)
\end{equation}
($d$ beeing the dimension of $M$; see e.g. \cite{BambergSternberg}-\cite{Feckoangl}).
In components (w.r.t. a local frame field $e_a$), it is defined as
\begin{equation} \label{defHodge2}
                 (*\alpha)_{a\dots b} := \frac{1}{p!}\alpha_{c\dots d} \ \omega^{c\dots d}_{ \ \ \ \ a\dots b} \ ,
                 \hskip 1cm
                 \omega^{c\dots d}_{ \ \ \ \ a\dots b} := g^{ce}\dots g^{df}\omega_{e\dots fa\dots b} \ ,
\end{equation}
where $\omega \equiv \omega_{g,o}$ stands for the \emph{metric volume form} on $(M,g,o)$.

The Hodge star $*$ is a remarkably effective tool for concise and coordinate-free expression
of important physics equations.
For example, the Maxwell equations may be written, on any (Lorentzian) spacetime, as
\begin{equation} \label{MaxwellMinkowski}
                 dF = 0 \ ,
                 \hskip 1cm
                 d* F = - J
\end{equation}
(here $J$ stands for $(d-1)$-form of current). Notice that no linear connection is needed for that
(no semicolons in local coordinates).

Recently, much research has been devoted to the study of physics in \emph{Galilean} and \emph{Carrollian} spacetimes,
see e.g. \cite{DuvGibHorZha2014}-\cite{HenneaxSalReb2021}
or a long lists of references in recent PhD theses \cite{GrassiePhD} and \cite{HansenPhD}.
Now if we wanted to use Hodge star operator on these spacetimes,
a serious problem would immediately arise:
There is \emph{no canonical} metric tensor on either Galilean or Carrollian spacetime.
And since metric tensor $g$ turns out to be key element in construction of the $*_g$,
the routine procedure fails, there.

\section{Galilean and Carrollian Hodge stars}
\label{galileiCarrollHodge}

From (\ref{defHodge2}) we see that the role of the metric tensor $g$
in construction of the ``mixed'' tensor $\omega^{c\dots d}_{ \ \ \ \ a\dots b}$,
i.e. in construction of the (standard) Hodge star, is twofold:
It enters the scene via
\vskip .1cm

- the \emph{metric} volume form $\omega_g$ and

- the co\emph{metric} $g^{-1}$ (for raising indices).

\vskip .1cm
\noindent
Concequently, \emph{if} the two objects, \emph{canonical} volume form and $\binom 20$-type
tensor field (for raising indices) were also available on Galilean/Carrollian spacetimes,
one could also construct $\omega^{c\dots d}_{ \ \ \ \ a\dots b}$ there,
i.e. potentially useful \emph{analogs} of the Hodge star were possible on \emph{Galilean} (\emph{Carrollian})
spacetime as well.

More generally, analogs of the standard Hodge star could be defined on Galilean and Carrollian spacetimes,
if the necessary mixed tensor field $\omega^{c\dots d}_{ \ \ \ \ a\dots b}$
could be constructed using \emph{canonical} tensor fields
\emph{available} there.

It turns out that such a construction is indeed possible.

Let us start with the fact (see Appendix \ref{galileicarrollmanifold}) that,
on \emph{both} Galilean and Carrollian spacetimes, \emph{both} (canonical) top-degree form and top-degree polyvector
\begin{equation} \label{galcartopformpolyvector}
                 \omega_{a \dots b} \ ,
                 \hskip 2cm
                 \tilde \omega^{a \dots b} \ ,
\end{equation}
are available, see (\ref{topdegreeomega}) and (\ref{topdegreeomegacomponents}).

So we have at least two different possibilities for constructing the needed \emph{mixed} tensor
$\omega^{a \dots b}_{ \ \ \ \ c\dots d}$: we
\vskip .2cm

- \emph{either} start from $\omega_{a \dots bc\dots d}$ %given in (\ref{topdegreeomegacomponents})
  and \emph{raise} (somehow) the group $(a\dots b)$,

- \emph{or} start from $\tilde \omega^{a \dots bc\dots d}$ %given also in (\ref{topdegreeomegacomponents})
   and \emph{lower} (somehow) the group $(c\dots d)$.

\vskip .2cm
\noindent
Since we start from \emph{canonical} tensors $\omega_{a \dots bc\dots d}$ and $\tilde \omega^{a \dots bc\dots d}$,
the resulting tensor $\omega^{a \dots b}_{ \ \ \ \ c\dots d}$ will be canonical as well
provided that the \emph{procedures} of raising and lowering indices will be canonical, too.

Now recall (see Sections (\ref{galileanmanifold}) and (\ref{carrollianmanifold}))
what canonical tensors (with the potential of raising and lowering indices) are available on the spacetimes of our interest:
\vskip .2cm

- Galilean: \hskip .3cm $(M, \xi, h), \quad h\in \binom 20, \quad \xi \in \binom 01
                        \quad \Rightarrow \quad k:= \xi \otimes \xi \in \binom 02$,

- Carrollian: $(M, \tilde \xi, \tilde h), \quad \tilde h\in \binom 02, \quad \tilde \xi \in \binom 10
              \quad \Rightarrow \quad \tilde k := \tilde \xi \otimes \tilde \xi \in \binom 20$.

\vskip .2cm
\noindent
So indeed, on both spacetimes of interest, there is a possibility to both raise and lower indices
in a \emph{canonical} way:
\vskip .2cm

- Galilean: \hskip .3cm \quad raising with $h\in \binom 20$, \quad lowering with $k \in \binom 02$,

- Carrollian: \quad raising with $\tilde k \in \binom 20$, \quad lowering with $\tilde h\in \binom 02$.

\vskip .2cm
\noindent
This means that there are, on \emph{both} spacetimes of interest, as many as two \emph{analogs of the Hodge star} operator
(in the sense of the canonical mapping of $p$-forms to $(d-p)$-forms on $d$-dimensional $M$),
namely those based on tensors $h^{ab}$/$\tilde h_{ab}$ and those based on tensors $k_{ab}$/$\tilde k^{ab}$.

In order to foresee some important features of the resulting operators, just recall
(see again Sections (\ref{galileanmanifold}) and (\ref{carrollianmanifold})) how (matrices of)
components of the necessary canonical tensors look like (in \emph{any adapted} frame/coframe):
\begin{eqnarray} %{rcl}
      \label{Galileimatrixh}
             \text{Galilean}  &:& h^{ab} \leftrightarrow \left( \begin{matrix}
                                                    h^{00} & h^{0i}  \\
                                                    h^{i0} & h^{ij}
                                                    \end{matrix}
                                             \right)
                             =
                                             \left( \begin{matrix}
                                                    0 & 0  \\
                                                    0 & \ \delta^{ij}
                                                    \end{matrix}
                                             \right),                                     \\
      \label{Galileimatrixk}
                         {}  &{}& k_{ab} \leftrightarrow \left( \begin{matrix}
                                                    k_{00} & k_{0i}  \\
                                                    k_{i0} & k_{ij}
                                                    \end{matrix}
                                             \right)
                             =
                                             \left( \begin{matrix}
                                                    1 & 0  \\
                                                    0 & 0
                                                    \end{matrix}
                                             \right).
\end{eqnarray}
\begin{eqnarray} %{rcl}
      \label{Carrollmatrixtildeh}
             \text{Carrollian}  &:& \tilde h_{ab} \leftrightarrow \left( \begin{matrix}
                                                    \tilde h_{00} & \tilde h_{0i}  \\
                                                    \tilde h_{i0} & \tilde h_{ij}
                                                    \end{matrix}
                                             \right)
                             =
                                             \left( \begin{matrix}
                                                    0 & 0  \\
                                                    0 & \ \delta_{ij}
                                                    \end{matrix}
                                             \right),                                      \\
      \label{Carrollmatrixtildek}
                         {}  &{}& \tilde k^{ab} \leftrightarrow \left( \begin{matrix}
                                                    \tilde k^{00} & \tilde k^{0i}  \\
                                                    \tilde k^{i0} & \tilde k^{ij}
                                                    \end{matrix}
                                             \right)
                             =
                                             \left( \begin{matrix}
                                                    1 & 0  \\
                                                    0 & 0
                                                    \end{matrix}
                                             \right).
\end{eqnarray}
Since the matrices $h^{ab}$ and $\tilde h_{ab}$ are typically much \emph{less} degenerate (their rank being $n$)
than the matrices $k_{ab}$ and $\tilde k^{ab}$ (rank $1$),
the procedure of raising/lowering indices using tensors $h^{ab}$/$\tilde h_{ab}$ is much less ``lossy''
than that using tensors $k_{ab}$/$\tilde k^{ab}$.
(Recall that raising and lowering indices is perfectly ``lossless'' in the standard Riemannian case.)
Put it differently, analogs of Hodge star based on $h^{ab}$/$\tilde h_{ab}$
are \emph{in general} expected to be \emph{more useful}
than those based on tensors $k_{ab}$/$\tilde k^{ab}$.

All the operators are computed in detail in Appendix \ref{fourhodgeanalogs} and the results are
summarized and discussed in Section \ref{formulasproperties}.

\section{Explicit formulae and some properties}
\label{formulasproperties}

\subsection{General explicit formulae}
\label{galileicarrollHodgeformula}

In both Galilean and Carrollian spacetimes, exactly as is the case in Lorentzian spacetime
(see e.g. \cite{Fecko3+1,FeckoAPS} and Sec.16.1 in \cite{Feckoangl}),
any $p$-form $\alpha$ may be uniquely written, w.r.t. (local) adapted coframe field $e^a = (e^0,e^i)$, as
\begin{equation} \label{alphadecomp}
                 \alpha = e^0 \wedge \hat s + \hat r \ ,
\end{equation}
where the two hatted forms, $(p-1)$-form $\hat s$ and $p$-form $\hat r$, respectively, are \emph{spatial},
meaning that they do not contain the (``temporal'') 1-form $e^0$.

When applying both versions of Hodge star operator $*$ onto $(1+n)$-decomposed presentation
(\ref{alphadecomp}) of $\alpha$, we get (in any dimension $d=1+n$) the following results
(see Appendix \ref{fourhodgeanalogs} for more details).

First, Hodge stars based on (rank $n$) tensors $h^{ab}$ and $\tilde h_{ab}$, respectively,
          act as follows:
\footnote{We display the results, just for comparison, together with the Lorentzian case,
          based on standard metric $g_{ab}$.
          Notice that all this is compatible with an
          %comes out easily from
          observation that, naively,
          Galilean limit ($c\to \infty$) of the Lorentzian case may be written as $e^0\gg e^i$ and for
          Carrollian limit ($c\to 0$) we have $e^0\ll e^i$.}
\begin{align} %{rcl}
              \label{starMinkowski1}
              * (e^0 \wedge \hat s + \hat r) =& \ e^0 \wedge \hat * \hat r + \hat * \hat \eta \hat s
              &   &\text{\emph{Lorentzian} Hodge star},   \\
              \label{starGalilei1}
              * (e^0 \wedge \hat s + \hat r) =& \ e^0 \wedge \hat * \hat r
              &   &\text{\emph{Galilean} Hodge star},  \\
              \label{starCarroll1}
              * (e^0 \wedge \hat s + \hat r) =& \hskip 1.75cm \hat * \hat \eta \hat s
              &   &\text{\emph{Carrollian} Hodge star}.
\end{align}
(see e.g. \cite{Fecko3+1,FeckoAPS} for (\ref{starMinkowski1}),
 (\ref{*onalphawithGalhab}) and (\ref{*onalphawithCartildehab})
 for (\ref{starGalilei1}) and (\ref{starCarroll1})).
Here $\hat *$ stands for standard ``Euclidean'' ($n$-dimensional) Hodge star (it only operates on spatial forms)
and $\hat \eta$ is the ``main automorhism'' on forms
(just a multiple by $(-1)^p$; see e.g. Sec.5.3 in \cite{Feckoangl}).

 Notice that this type of
Galilean $*$ vanishes for $p=d=1+n$ (top-degree forms) and the Carrollian $*$ vanishes for $p=0$ (functions).

Second, there are Hodge stars based on (rank $1$) tensors
$k_{ab}$ (for Galilean case) and $\tilde k^{ab}$ (for Carrollian case).
Their action is displayed in
(\ref{*Galkabp<n}) - (\ref{*Galkabp=n+1}) for Galilean case and
(\ref{*Cartildekabp=0}) - (\ref{*Cartildekabp=2dots}) for Carrollian case.
The comparison shows the following:

\vskip .2cm
\noindent
\emph{Galilean}: The ``Hodge star'' $*$ based on $k_{ab}$

- vanishes for all $p<n$ (unlike those given by (\ref{starGalilei1})),

- coincides for $p=n$ with (\ref{starGalilei1}),

- does not vanish for $p=n+1$ (unlike the one given by (\ref{starGalilei1})).

\vskip .2cm
\noindent
\emph{Carrollian}: The ``Hodge star'' $*$ based on $\tilde k^{ab}$

- vanishes for all $p>1$ (unlike those given by (\ref{starCarroll1})),

- coincides for $p=1$ with (\ref{starCarroll1}),

- does not vanish for $p=0$ (unlike the one given by (\ref{starCarroll1})).

\vskip .2cm
Now, it is clear that \emph{zero} operators from $p$-forms to $q$-forms are,
although being formally canonical, not very interesting,
notably if there are also \emph{non-zero} canonical operators available in the same situation.

So, from this point of view, we can conclude that we should
\vskip .2cm
\noindent

- \emph{in general} choose (\ref{starGalilei1}) and (\ref{starCarroll1}) ,

\vskip .2cm
\noindent
i.e. star operators based on (less degenerate) tensors $h^{ab}$ and $\tilde h_{ab}$,
the \emph{only exceptions} being

\vskip .2cm
\noindent

- \emph{Galilean}: \hskip .3cm$p=d=1+n$, when (\ref{*Galkabp=n+1}) is more useful (= non-zero) than (\ref{starGalilei1}),

- \emph{Carrollian}: $p=0$, when (\ref{*Cartildekabp=0}) is more useful (= non-zero) than (\ref{starCarroll1}).

\vskip .2cm
\noindent
So the two \emph{exceptional} cases (which differ from (\ref{starGalilei1}) and (\ref{starCarroll1})) read
\begin{align} %{rcl}
      \label{*Galkabp=n+1_2}
      *(e^0 \wedge \hat \omega )    =& \ 1                        &   \text{\emph{Galilean} Hodge star}& &  \text{for} \ \ &p=1+n, \\
      \label{*Cartildekabp=0_2}
      *1                            =& \ e^0 \wedge \hat \omega   & \text{\emph{Carrollian} Hodge star}& &  \text{for} \ \ &p=0.
\end{align}
For even more explicit expressions of (\ref{starMinkowski1}) - (\ref{starCarroll1})
and (\ref{*Galkabp=n+1_2}) - (\ref{*Cartildekabp=0_2}) for the usual $1+3$ (rather than $1+n$) case,
see Appendix \ref{app:galileiexplformulae}.

For a completely different approach (based on the concept of \emph{intertwining ope\-rator})
to finding formulas (\ref{*MinkOmega0}) - (\ref{*CarOmega4}) from Appendix \ref{app:galileiexplformulae},
which also provides an interesting additional insight, see paper \cite{FeckoGalCaroperators}.

\subsection{Basic properties}
\label{galileiHodgeproperties}

Exactly as is the case for the standard Hodge star, both new Hodge stars map linearly
$p$-forms on $(d-p)$-forms and vice versa,
\begin{equation} \label{pton-pviceversa}
          \Omega^p  \underset * {\overset * \rightleftarrows}   \Omega^{d-p} \ .
\end{equation}
A straightforward consequence of (\ref{starGalilei1}) and (\ref{starCarroll1})
          is a striking feature of the new Hodge stars - they both square to \emph{zero}
\footnote{Except for the ``extremal'' degrees $p=0$ and $p=d$ when, as is mentioned
          at the end of Section \ref{galileicarrollHodgeformula},
          we prefer (\ref{*Galkabp=n+1}) to (\ref{starGalilei1}) and
          (\ref{*Cartildekabp=0}) to (\ref{starCarroll1}).
          In these exceptional cases it squares to (plus or minus) the unit operator, exactly as we are accustomed to
          in the standard case.},
\begin{equation} \label{iteratedvanishes}
                 ** = 0 \ ,
                 \hskip 1cm
                 ** : \Omega^p \to \Omega^p \ ,
                 \hskip 1cm
                 0 \neq p \neq   d \ .
\end{equation}
This is in sharp contrast with the standard Hodge operator $*_g$ from (\ref{defHodge2}) which, as is well-known,
squares to (plus or minus) the \emph{unit} operator,
\begin{equation} \label{*gsquared}
                 *_g*_g = \pm \hat 1 \ ,
                 \hskip 1cm
                 *_g*_g : \Omega^p \to \Omega^p \ ,
                 \hskip 1cm
                 \text{for \emph{any}} \ p \ .
\end{equation}
So, contrary to the standard case,
neither the \emph{Galilean}-Hodge star
nor the \emph{Carrollian}-Hodge star
are \emph{isomorphisms} between the two spaces of equal dimensions
(spaces of $p$-forms and $(d-p)$-forms, respectively; exceptions are $p=0$ and $p=d$).
Their (non-zero) kernels are
\begin{align} %{rcl}
              \label{starGalilei10}
              * (e^0 \wedge \hat s) =& \ 0
              &   &\text{\emph{Galilean} case}
              &   &\Deg \hat s \neq n,           \\
              \label{starCarroll10}
              * \ \hat r =& \ 0
              &   &\text{\emph{Carrollian} case}
              &   &\Deg \hat r \neq 0.
\end{align}
One can also rephrase the statement as that we can \emph{no longer} speak of
Galilean-Hodge \emph{duality} as well as
Carrollian-Hodge \emph{duality} (with the above mentioned exceptions).

That is the price to pay for replacing
\emph{Lorentzian} spacetime with
\emph{Galilean} and
\emph{Carrollian} spacetimes,
while still insisting that the concept of ``Hodge star operator'' remains \emph{canonical}.

\subsection{Behaviour with respect to pull-back}
\label{pullback}

From the component expression (\ref{defHodge2}) for the \emph{standard} Hodge star operator
one can see that the corresponding component-free expression  reads
\begin{equation} %{rcl}
              \label{*compfree}
              *_g \alpha \ \sim \ C \dots C (g^{-1} \otimes \dots \otimes g^{-1} \otimes \omega_g \otimes \alpha) \ ,
\end{equation}
where $g^{-1}$ denotes cometric and $C$ stands for contraction.

Now, let
\begin{equation} %{rcl}
              \label{fMtoM}
              f:M\to M
\end{equation}
be a diffeomorphism of $M$. Then, taking into account
\begin{equation} %{rcl}
              \label{f*ofcometricandvolform}
              f^*(g^{-1}) = (f^*g)^{-1} \ ,
              \hskip 1cm
              f^*\omega_g = \omega_{f^*g} \ ,
\end{equation}
we get a well-known (and useful) property of the (standard) Hodge star:
\begin{equation} %{rcl}
              \label{f^**_g}
              f^*(*_g\alpha) = *_{f^*g}(f^*\alpha) \ ,
              \hskip 1cm \text{i.e.} \hskip 1cm
              f^* \circ *_g = *_{f^*g} \circ f^* \ .
\end{equation}
Now, consider replacing the (peudo)Riemannian manifold $(M,g)$ with a Galilean manifold $(M,h,\xi)$
or a Carrollian manifold $(M,\tilde h,\tilde \xi)$.
The \emph{structure} (\ref{*compfree}) of the Hodge star remains the same,
we just make substitutions of tensor fields
(depending on particular version, out of four possibilities, of the star,
 discussed in Section \ref{galileiCarrollHodge}),
\begin{align} %{rcl}
              \label{substitGal1}
              &\text{\emph{Galilean} case} \ :
              &   &1. & g^{-1} &\mapsto h                                  \ ,
              &   \omega_g &\mapsto \omega_{h,\xi}                         \ ,       \\
              \label{substitGal2}
              &{}
              &   &2. & g^{-1} &\mapsto k                                  \ ,
              &   \omega_g &\mapsto {\tilde \omega}_{h,\xi}                \ .       \\
              \label{substitCar1}
              &\text{\emph{Carrollian} case} \ :
              &   &1. & g^{-1} &\mapsto \tilde h                           \ ,
              &   \omega_g &\mapsto \tilde \omega_{\tilde h,\tilde \xi}    \ ,       \\
              \label{substitCar2}
              &{}
              &   &2. & g^{-1} &\mapsto \tilde k                           \ ,
              &   \omega_g &\mapsto \omega_{\tilde h,\tilde \xi}           \ .
\end{align}
(where $\omega_{h,\xi}$, $\tilde \omega_{h,\xi}$, $\omega_{\tilde h,\tilde \xi}$ and $\tilde \omega_{\tilde h,\tilde \xi}$
 are defined in (\ref{topdegreeomega})
 and $k$ and $\tilde k$ is introduced in Section \ref{galileiCarrollHodge}).
Then, taking into account
\begin{equation} %{rcl}
              \label{f*ofvolforms}
              f^*\omega_{h,\xi} = \omega_{f^*h,f^*\xi}
              \hskip 1cm
              f^*\tilde \omega_{\tilde h,\tilde \xi} = \tilde \omega_{f^*\tilde h,f^*\tilde \xi} \ ,
\end{equation}
(and similarly for all other ``derived'' canonical tensor fields mentioned above)
we get corresponding useful property of the ``new'' Hodge stars:
\begin{equation} %{rcl}
              \label{f^**GalCar}
              f^* \circ *_{h,\xi} = *_{f^*h,f^*\xi} \circ f^*
              \hskip 1cm \text{and} \hskip 1cm
              f^* \circ *_{\tilde h,\tilde \xi} = *_{f^*\tilde h,f^*\tilde \xi} \circ f^* \ .
\end{equation}
(where $*_{h,\xi}$ denotes \emph{any} of the two Galilean stars and similarly $*_{\tilde h,\tilde \xi}$
 denotes \emph{any} of the two Carrolian stars).
In particular, for \emph{structure-preserving} diffeomorphisms, i.e. such that
\begin{align} %{rcl}
              \label{structpresg}
              &\text{in \emph{Lorentzian} case},
              &   f^*g &= \ g,                           \\
              &\text{in \emph{Galilean} case},
              &   f^*h &= \ h,  \hskip .4cm
                 f^*\xi = \ \xi,                         \\
              &\text{in \emph{Carrollian} case},
              &   f^*\tilde h &= \ \tilde h,  \hskip .4cm
                 f^*\tilde \xi = \ \tilde \xi.
\end{align}
holds the \emph{corresponding} Hodge star \emph{commutes} with pull-back,
\begin{equation} %{rcl}
              \label{f^**=*f^*all}
              f^* \circ * = * \circ f^*.
\end{equation}
So, as an example, for the simplest versions of the two non-lorentzian spacetimes,
\emph{Galilei} and \emph{Carroll} spacetimes,
the corresponding star operators commute with (translations, rotations plus)
\begin{alignat}{5} %{rcl}
 \label{tgalcarflat}
 &\text{Galilei boost}\colon \ && t' = t,\qquad &&
 \text{Carroll boost}\colon \ && t' = t + \mathbf v \cdot \mathbf r \ ,& \\
 \label{rgalcarflat}
&&& \mathbf r' = \mathbf r + \mathbf v t ,\qquad &&&&
 \mathbf r' = \mathbf r \ .&
\end{alignat}
(As for the Carroll boost, see e.g. (9) in \cite{LevyLeblond}, II.10 in \cite{DuvGibHorZha2014}
 or Appendix \ref{carrollianmanifold}.)
Put it differently, the new operators happen to be \emph{invariant} w.r.t. Galilei and Carroll transformations, respectively.

More generally, the new ``Hodge stars'' are invariant w.r.t. \emph{local Galilean} and \emph{local Carrollian}
transformations, respectively, in the same way as, say, the standard Hodge star
on a Lorentzian spacetime is known to be invariant w.r.t. \emph{local Lorentzian} transformations.

This property makes the Galilean and Carrollian Hodge star operators potentially interesting
from the point of view of writing important \emph{physics equations}
in terms of \emph{differential forms} on corresponding \emph{spacetimes} (see an example in Section \ref{galileicarrollmaxwell}).

\section{An application: Electrodynamics}
\label{galileicarrollelectrodynamics}

As a simple application of Galilean/Carrollian Hodge star operator in physics,
let us have just a brief look at Galilei and Carroll \emph{electrodynamics}.
In these two versions of electrodynamics (particular limits of the standard Minkowski one,
see e.g. \cite{DuvGibHorZha2014,BellacLevyLeblond}),
the corresponding equations of motion for the fields $\mathbf E$ and $\mathbf B$
are Galilei/Carroll (rather than Poincar\'e) invariant.

\subsection{Standard electrodynamics}
\label{lorentzmaxwell}

In \emph{Minkowski} spacetime,
the (source-less, for simplicity) Maxwell equations
\begin{eqnarray} %{rcl}
      \label{d*F=0Mina}
             \Div \mathbf E                          &=& \ 0 \ , \\
      \label{d*F=0Minb}
             \curl \mathbf B - \partial_t \mathbf E  &=& \ 0 \ , \\
      \label{dF=0Mina}
             \curl \mathbf E + \partial_t \mathbf B  &=& \ 0 \ , \\
      \label{dF=0Minb}
             \Div \mathbf B                          &=& \ 0 \ ,
\end{eqnarray}
may be neatly written in terms of 2-form of electromagnetic field,
\begin{equation} \label{F2formelmagMink}
                 F = dt \wedge \mathbf E \cdot d\mathbf r - \mathbf B \cdot d\mathbf S \ ,
\end{equation}
as follows:
\begin{eqnarray} %{rcl}
      \label{d*F=0Mink}
             d* F
                  &=& \ 0 \ , \hskip 2cm * = *_\eta \ \ \ \text{here}  , \\
      \label{dF=0Mink}
             dF
                  &=& \ 0 \ .
\end{eqnarray}
As already mentioned in Section \ref{pullback},
just because of properties of ($d$ and) $*_\eta$,
this way of presentation of the Maxwell equations makes their \emph{Poincar\'e} invariance evident.

In more detail, the assignment reads
\begin{align} %{rcl}
              \label{gauss}
                    d*F             &= 0   & &\leftrightarrow & \Div \mathbf E                          &= 0 \ ,       \\
              \label{ampere}
                                    &{}    & &{}              & \curl \mathbf B - \partial_t \mathbf E  &= 0 \ ,       \\
              \label{faraday}
                     dF             &= 0   & &\leftrightarrow & \curl \mathbf E + \partial_t \mathbf B  &= 0 \ ,    \\
              \label{nomonopole}
                                    &{}    & &{}              & \Div \mathbf B                          &= 0
\end{align}
(see, e.g., Section 16.1 in \cite{Feckoangl}). Notice that each spacetime equation on the l.h.s.
corresponds to as many as \emph{two} spatial equations on the r.h.s.

\subsection{Galilei and Carroll electrodynamics}
\label{galileicarrollmaxwell}

Now it looks natural, in an effort to switch to Galilei (Carroll) electrodynamics,
to just repeat the complete story with \emph{Galilei} (\emph{Carroll}) Hodge star ope\-ra\-tor.
So again, introduce (see (\ref{Omega2})) the 2-form of electromagnetic field exactly as in (\ref{F2formelmagMink})
and write down equations of motion \`{a} la (\ref{d*F=0Mink}) and (\ref{dF=0Mink}),
the only change being the replacement of the Minkowski Hodge star with the Galilei (Carroll) one
(so with $*$ from (\ref{starGalilei1}) or (\ref{starCarroll1}); more explicitly from (\ref{*GalOmega2}) or (\ref{*CarOmega2})):
\begin{eqnarray} %{rcl}
      \label{d*F=0Gal}
             d* F
                  &=& \ 0 \ ,  \hskip 2cm * = *_{h,\xi} \ \text{or} \  *_{\tilde h,\tilde \xi} \ , \ \text{here} \ , \\
      \label{dF=0Gal}
             dF
                  &=& \ 0 \ .
\end{eqnarray}
Then it is clear that, again just because of the properties of $d$ and $*$,
this is
a~system of 1-st order partial differential equations
for the fields $\mathbf E$ and $\mathbf B$,
which is \emph{Galilei} or \emph{Carroll} (rather than Poincar\'e) \emph{invariant}.

\vskip .2cm
\noindent
All the items listed above provide a promising signal that equations (\ref{d*F=0Gal}) - (\ref{dF=0Gal})
are probably closely associated with the desired Galilei (Carroll) electrodynamics.

How the system (\ref{d*F=0Gal}) - (\ref{dF=0Gal}) actually looks like in terms of the fields $\mathbf E$ and $\mathbf B$?

First, it is clear that (\ref{dF=0Gal}) looks \emph{the same} for all three cases
(standard Minkowski, Galilei as well as Carroll), namely it always corresponds to
(\ref{faraday}) and (\ref{nomonopole}). So, both the Faraday's law and nonexistence of magnetic monopoles
hold in all three versions of electrodynamics.

What really makes a difference is equation (\ref{d*F=0Gal}).
When (\ref{*GalOmega2}) and (\ref{*CarOmega2}) - rather than (\ref{*MinkOmega2}) - is used for computation of $*$,
we get
\begin{align} %{rcl}
              \label{amperegal}
                    d*F             &= 0   & &\leftrightarrow & \curl \mathbf B         &= 0                          & &\text{in Galilei case} \ ,     \\
              \label{amperecar}
                    d*F             &= 0   & &\leftrightarrow & \partial_t \mathbf E    &= 0 \ , \ \Div \mathbf E = 0 & &\text{in Carroll case} \ .
\end{align}
So, what we get when \emph{all} spatial equations, resulting from
\begin{equation} \label{d*F=0dF=0}
                 d*F = 0 \ , \hskip 1.5 cm dF = 0 \ ,
\end{equation}
are displayed side by side for \emph{all three} cases is
\begin{align} %{rcl}
              %\label{LorGalCar}
                    \text{Minkowski}&  &  \text{Galilei}&    &  \text{Carroll}&     \\
              \label{Gaussall3}
                    \Div \mathbf E                          &= 0 \ , &                                         &{}      &                   \Div \mathbf E        &= 0 \ , \\
              \label{Ampereall3}
                    \curl \mathbf B - \partial_t \mathbf E  &= 0 \ , & \curl \mathbf B \hskip 1.1cm            &= 0 \ , &                   \partial_t \mathbf E  &= 0 \ , \\
              \label{Faradayall3}
                    \curl \mathbf E + \partial_t \mathbf B  &= 0 \ , & \curl \mathbf E + \partial_t \mathbf B  &= 0 \ , & \curl \mathbf E + \partial_t \mathbf B  &= 0 \ , \\
               \label{nomonopoleall3}
                    \Div \mathbf B                          &= 0 \ , & \Div \mathbf B                          &= 0 \ , & \Div \mathbf B                          &= 0 \ .
\end{align}
Checking equations in Galilei and Carroll columns versus \emph{standard} Maxwell equations (left column) shows that there are
some objects \emph{missing} in Galilei as well as Carroll versions of ``Maxwell equations'' (\ref{d*F=0dF=0}), namely
\vskip .2cm

- \emph{displacement current} $\partial_t \mathbf E$ in Amp\`{e}re's law in the \emph{Galilei} case,

- ``\emph{Amp\`{e}re} term'' $\curl \mathbf B$ in Amp\`{e}re's law in the \emph{Carroll} case,

- equation (\ref{Gaussall3}), i.e. \emph{Gauss's law}, in the \emph{Galilei} case.

\vskip .2cm
\noindent
Now, while the absence of the displacement current in Galilei electrodynamics and
the ``basic Amp\`{e}re term'' in Carroll electrodynamics
are well-known (so expected and desired) facts
(namely in the so-called ``magnetic limit'' version of Galilei electrodynamics, see \cite{BellacLevyLeblond},
 and ``electric limit'' version of Carroll electrodynamics, see \cite{DuvGibHorZha2014}),
the Gauss' law is usually present in all three versions of electrodynamics.
In particular, the Galilei system usually contains the equations from the central column
plus the equation (\ref{gauss}).
(The fact that the Gauss' law is not present in the spacetime differential forms version
 of the Galilei electrodynamics needs further study.
 We \emph{do get}, by the way, the Gauss' law with the help of Galilei \emph{connection}.)

So, to summarize, (\ref{d*F=0dF=0}) \emph{do} represent \emph{correct} equations
in \emph{all three} versions of electrodynamics,
albeit they do not provide, in the Galilei case, the whole story.

\vskip .5cm
\noindent
[Technically, the reason why just a \emph{single} equation (\ref{amperegal}) corresponds
          to (\ref{d*F=0Gal})
          (so that, at the end of the day, one equation is missing),
          contrary to \emph{two} equations (\ref{gauss}) and (\ref{ampere}) in Minkowski case,
          looks to be simple: The Galilei star is no longer an isomorphism on 2-forms,
          it \emph{kills} once and for all the ``electric'' part of $F$ and then the action of $d$ only produces a single term.
          Notice, however, that neither the \emph{Carroll} star is an isomorphism on 2-forms
          (it kills once and for all the ``magnetic'' part of $F$). And here action of $d$ produces \emph{two} terms.
          So the number of resulting equations \emph{alone} actually depends on more subtle \emph{details}
          of actions of the two star operators, rather than on their general common property of not being isomorphism.]
\vskip .5cm
\noindent

\section{Summary and conclusions}
\label{summaryconclusions}

The Hodge star operator belongs to essential tools in applications of differential forms on (pseudo)Riemannian manifolds.
In particular, this is true for standard Lorentzian spacetimes.

However, Galilean and Carrollian spacetimes are not (pseudo)Riemannian mani\-folds.
There is no (distinguished, non-degenerate) metric tensor available on them.
Consequently, there is no standard ``full-fledged'' Hodge star operator on them.

Nevertheless, it turns out that one can \emph{mimic} the usual construction of the Hodge star
using specific (well-known) canonical tensors available.
Since algebra of canonical tensors depends on particular spacetime,
particular constructions and the resulting ``Hodge stars'' depend on particular spacetime as well.

The new ``Hodge stars'' retain some important properties of the standard one, while losing others.

In particular, they remain to be \emph{invariant} w.r.t. structure preserving diffeomorphisms
(in particular w.r.t. Galilei or Carroll transformations, respectively).
More generally, w.r.t. \emph{local} Galilean and \emph{local} Carrollian transformations, respectively.
This makes them potentially interesting from the point of view of writing \emph{physics equations}
in terms of differential forms on Galilean and Carrollian spacetimes.

Both new ``Hodge stars'', however, cease to be linear \emph{isomorphisms}
(in general). One can no longer speak of Hodge \emph{duality} (modulo exceptions).

In the three spacetimes under consideration (Lorentzian, Galilean and Carrollian) one can express differential forms
in terms of a \emph{pair} of \emph{spatial} forms. Then the three Hodge star operators act according to
(\ref{starMinkowski1}) - (\ref{starCarroll1}) or
(\ref{*Galkabp=n+1_2}) - (\ref{*Cartildekabp=0_2}).

For example, on Minkowski, Galilei and Carroll $(1+3$)-dimensional spacetimes,
the action on the 2-form $F$ of electromagnetic field reads
\begin{eqnarray} %{rcl}
      \label{MinstaronF}
             *_M (dt \wedge \mathbf E \cdot d\mathbf r - \mathbf B \cdot d\mathbf S)
                  &=& dt \wedge (-\mathbf B) \cdot d\mathbf r - \mathbf E \cdot d\mathbf S \ , \\
      \label{GalstaronF}
             *_G (dt \wedge \mathbf E \cdot d\mathbf r - \mathbf B \cdot d\mathbf S)
                  &=& dt \wedge (-\mathbf B) \cdot d\mathbf r  \ ,                             \\
      \label{CarstaronF}
             *_C (dt \wedge \mathbf E \cdot d\mathbf r - \mathbf B \cdot d\mathbf S)
                  &=& \hskip 2.5cm                - \mathbf E \cdot d\mathbf S \ .
\end{eqnarray}
Consequently, single (universal) \emph{manifestly invariant} equation on spacetime
\begin{equation} \label{d*F=0}
                 d*F = 0 \ ,
\end{equation}
has substantially different 3-dimensional content
(expression in terms of fields $\mathbf E$ and $\mathbf B$),
depending on \emph{which} particular Hodge star operator ($*_M$, $*_G$ or $*_C$)
is used
(see (\ref{Gaussall3}) and (\ref{Ampereall3}).

\appendix

\section{Appendices}
\label{appendices}

\subsection{Galilean and Carrollian manifolds}
\label{galileicarrollmanifold}

Here, for convenience of the reader, basic definitions and facts are collected.
In order to learn more details, one should consult references at the end of the paper,
e.g.,
\cite{DuvGibHorZha2014}-\cite{HenneaxSalReb2021}
and
\cite{Trautman1963}-\cite{LevyLeblond}.

\subsubsection{Galilean manifold}
\label{galileanmanifold}

Oriented Galilean manifold is a 4-tuple $(M, \xi, h,o)$, where

\vskip .1cm

- $(M,o)$ is an $d$-dimensional oriented manifold, $d=1+n$,

- $\xi$ is an everywhere non-zero covector (i.e. a $\binom01$-tensor) field on $M$,

- $h$ is an everywhere rank-$n$ symmetric type-$\binom20$-tensor field on $M$,

- such that $h (\xi, \ \cdot \ ) =0$.
\vskip .1cm
\noindent
We call a (local) right-handed frame field $e_a = (e_0,e_i)$ and the (dual) coframe field $e^a = (e^0,e^i)$, $i=1,\dots ,n$
on $(M, \xi, h,o)$ \emph{adapted} (or distinguished) if
\vskip .1cm

- $e^0 = \xi$,

- $h = \delta^{ij}e_i \otimes e_j$,

\vskip .1cm
\noindent
so that, in (any) adapted frame field, the components of the two tensor fields have "canonical form"
\begin{equation} \label{galilei01a20tensorsmatrices}
                 \xi_a \leftrightarrow \left( \begin{matrix}
                                                    \xi_0  \\
                                                    \xi_i
                                                    \end{matrix}
                                             \right)
                             =
                                             \left( \begin{matrix}
                                                    1  \\
                                                    0
                                                    \end{matrix}
                                             \right) \ ,
             \hskip 1cm
                 h^{ab} \leftrightarrow \left( \begin{matrix}
                                                    h^{00} & h^{0i}  \\
                                                    h^{i0} & h^{ij}
                                                    \end{matrix}
                                             \right)
                             =
                                             \left( \begin{matrix}
                                                    0 & 0  \\
                                                    0 & \ \delta^{ij}
                                                    \end{matrix}
                                             \right)
                             =
                                             \left( \begin{matrix}
                                                    0 & 0  \\
                                                    0 & \ \mathbb I
                                                    \end{matrix}
                                             \right) \ .
\end{equation}
The (point-dependent) change-of-basis matrix $A$ between any pair $\hat e_a$, $e_a$ of adapted frame fields, given by $\hat e_a = A_a^be_b$,
has the structure
\begin{equation} \label{Galileichange-of-basismatrix}
                      A^b_a \leftrightarrow \left( \begin{matrix}
                                                    A^0_0 & A^0_i  \\
                                                    A^i_0 & A^i_j
                                                    \end{matrix}
                                             \right)
                             =
                                             \left( \begin{matrix}
                                                    1   & 0  \\
                                                    v^i & \ R^i_j
                                                    \end{matrix}
                                             \right)
                             \hskip .5cm \text{i.e.} \hskip .5cm
                           A \leftrightarrow \left( \begin{matrix}
                                                    1   & 0  \\
                                                    v & \ R
                                                    \end{matrix}
                                             \right)
\end{equation}
where $R$ is an $n$-dimensional rotation matrix.

In each point, such matrices form a Lie group $G$, subgroup of $GL(d,\mathbb R)$,
the (homogeneous) \emph{Galilei group} ($R$ parametrizes rotations and $v$ boosts, respectively).

\subsubsection{Carrollian manifold}
\label{carrollianmanifold}

Oriented Carrollian manifold is a 4-tuple $(M, \tilde \xi, \tilde h,o)$, where

\vskip .1cm

- $(M,o)$ is an $d$-dimensional oriented manifold, $d=1+n$,

- $\tilde \xi$ is an everywhere non-zero vector (i.e. a $\binom10$-tensor) field on $M$,

- $\tilde h$ is an everywhere rank-$n$ symmetric type-$\binom02$-tensor field on $M$,

- such that $\tilde h (\tilde \xi, \ \cdot \ ) =0$.
\vskip .1cm
\noindent
We call a (local) right-handed frame field $e_a = (e_0,e_i)$ and the (dual) coframe field $e^a = (e^0,e^i)$,
$i=1,\dots ,n$ on $(M, \tilde \xi, \tilde h,o)$ \emph{adapted} (or distinguished) if
\vskip .1cm

- $e_0 = \tilde \xi$,

- $\tilde h = \delta_{ij}e^i \otimes e^j$,

\vskip .1cm
\noindent
so that, in (any) adapted frame field, the components of the two tensors have "canonical form"
\begin{equation} \label{carroll10a02tensorsmatrices}
                 \xi^a \leftrightarrow \left( \begin{matrix}
                                                    \xi^0  \\
                                                    \xi^i
                                                    \end{matrix}
                                             \right)
                             =
                                             \left( \begin{matrix}
                                                    1  \\
                                                    0
                                                    \end{matrix}
                                             \right) \ ,
             \hskip 1cm
                 h_{ab} \leftrightarrow \left( \begin{matrix}
                                                    h_{00} & h_{0i}  \\
                                                    h_{i0} & h_{ij}
                                                    \end{matrix}
                                             \right)
                             =
                                             \left( \begin{matrix}
                                                    0 & 0  \\
                                                    0 & \ \delta_{ij}
                                                    \end{matrix}
                                             \right)
                             =
                                             \left( \begin{matrix}
                                                    0 & 0  \\
                                                    0 & \ \mathbb I
                                                    \end{matrix}
                                             \right)
\end{equation}
The (point-dependent) change-of-basis matrix $A$ between any pair $\hat e_a$, $e_a$
of adapted local frame fields, given by $\hat e_a = A_a^be_b$,
has the structure
\begin{equation} \label{Carrollchange-of-basismatrix}
                      A^b_a \leftrightarrow \left( \begin{matrix}
                                                    A^0_0 & A^0_i  \\
                                                    A^i_0 & A^i_j
                                                    \end{matrix}
                                             \right)
                             =
                                             \left( \begin{matrix}
                                                    1   & v_i  \\
                                                    0 & \ R^i_j
                                                    \end{matrix}
                                             \right) \ ,
                             \hskip .5cm \text{i.e.} \hskip .5cm
                           A \leftrightarrow \left( \begin{matrix}
                                                    1   & v^T  \\
                                                    0 & \ R
                                                    \end{matrix}
                                             \right)
\end{equation}
where $R$ is a $n$-dimensional rotation matrix.

In each point, such matrices form a Lie group $G$, subgroup of $GL(d,\mathbb R)$,
the (homogeneous) \emph{Carroll group} ($R$ parametrizes rotations and $v$ boosts, respectively).

\subsubsection{Top-degree forms and polyvectors}
\label{topdegformpolyvector}

Consider, on a patch $\mathcal O$ of a $(1+n)$-dimensional manifold $M$,
\emph{arbitrary local} frame field $e_a=(e_0,e_i)$ and the dual coframe field $e^a = (e^0,e^i)$.
In terms of the frame/coframe fields, we introduce (local, everywhere non-vanishing)
top-degree form $\omega (e)$ and top-degree polyvector $\tilde \omega (e)$
\begin{equation} \label{topdegreeomega}
                 \omega (e) := e^0\wedge e^1 \wedge \dots \wedge e^n \ ,
                 \hskip 1.5cm
                 \tilde \omega (e) := e_0\wedge e_1 \wedge \dots \wedge e_n \ .
\end{equation}
Introduce, similarly, a primed frame/coframe on primed $\mathcal O'$.
Then, on the overlap $\mathcal O \cap \mathcal O'$, one has
\begin{equation} \label{changeof(co)frame}
                 e'_a = A_a^be_b \ ,
                 \hskip 1cm
                 e'^a = (A^{-1})^a_be^b
\end{equation}
(where $A\equiv A(x)$ is a unique position-dependent regular matrix) and we standardly get
\begin{equation} \label{topdegreeomeganew}
                 \omega (e') = (\det A)^{-1} \ \omega (e) \ ,
                 \hskip 1cm
                 \tilde \omega (e') = (\det A) \ \tilde \omega (e) \ .
\end{equation}
Now consider restriction of the above situation to
\vskip .1cm

- oriented \emph{Galilean/Carrollian} manifold,

- \emph{adapted} frame/coframe fields.

\vskip .1cm
\noindent
Then the change-of-basis matrices $A$ have the form
(\ref{Galileichange-of-basismatrix}) or (\ref{Carrollchange-of-basismatrix}),
depending on whether we speak of Galilean or Carrollian manifold.
In both cases, however, we have clearly
\begin{equation} \label{detA=1}
                 \det A = 1
\end{equation}
and, therefore, on the overlap $\mathcal O \cap \mathcal O'$,
\begin{equation} \label{topdegreeomeganewequalsold}
                 \omega (e') = \omega (e),
                 \hskip 1cm
                 \tilde \omega (e') =  \tilde \omega (e).
\end{equation}
This means that both objects, $\omega$ and $\tilde \omega$,
are well-defined on the union $\mathcal O \cup \mathcal O'$ and, consequently, that
\emph{canonical global} objects are actually defined in terms of (local) expressions (\ref{topdegreeomega}).

Put another way,
\vskip .1cm

- on \emph{both} Galilean \emph{and} Carrollian (oriented) manifolds \ ,

- \emph{both} canonical \emph{volume form} and canonical \emph{top-degree polyvector} exist.
\footnote{It is clear, that \emph{any constant multiples} of the tensor fields discussed above are canonical as well.}

\vskip .1cm
\noindent
In components (w.r.t. arbitrary local \emph{adapted} frame/coframe fields) this may be written
in terms of the Levi-Civita symbols as
\begin{equation} \label{topdegreeomegacomponents}
                 \omega_{a \dots b} = \epsilon_{a \dots b} \ ,
                 \hskip 2cm
                 {\tilde \omega}^{a \dots b} = \epsilon^{a \dots b} \ .
\end{equation}

\subsection{Details of computation of the four analogs of Hodge star}
\label{fourhodgeanalogs}

Here, detailed calculations of (analogs of) Hodge star operators,
leading to results collected in Section \ref{galileicarrollHodgeformula}, are shown.

Component definition (\ref{defHodge2}) is equivalent to an expression for action of $*$ on
\emph{basis} $p$-forms (on $d$-dimensional manifold)
\begin{equation} \label{*nabaze}
                  *(e^a \wedge \dots \wedge e^b) = \frac{1}{(d-p)!}\omega^{a \dots b}_{\hskip .6cm c \dots d} \ e^c \wedge \dots \wedge e^d.
\end{equation}

What we need to calculate from (\ref{*nabaze}) is action of $*$ on a \emph{general decomposed} $p$-form, i.e.
\begin{equation} \label{starondecomposedalpha}
                 *\alpha = *(e^0 \wedge \hat s + \hat r) = \ ?
                 \hskip 1cm
                 \hat s = \hat s_{p-1} \ , \ \hat r = \hat r_p
\end{equation}
In \emph{Lorentzian} case, the answer reads
\begin{equation} \label{starondecompalphaMink}
                  *(e^0 \wedge \hat s + \hat r) = e^0 \wedge \hat * \hat r + \hat * \hat \eta \hat s
\end{equation}
(see e.g. \cite{Fecko3+1,FeckoAPS} or, in the simplest case, Section 16.1. in \cite{Feckoangl}),
where $\hat *$ is the Euclidean Hodge star and $\hat \eta$ is the main automorphism on forms, i.e. just multiplication by $(-1)^p$.
What we want here are similar formulas for all four new cases,
Galilean versus Carrollian case and $h^{ab}$/$\tilde h_{ab}$ versus $k_{ab}$/$\tilde k^{ab}$ case.

Because of specific features of the tensors $h^{ab}, \tilde h_{ab},k_{ab},\tilde k^{ab}$,
it is convenient to discuss the two parts ($e^0 \wedge \hat s$ and $\hat r$) separately,
i.e. to compute
\begin{equation} \label{staronmixedandspatial}
                 *(e^0 \wedge e^i \wedge \dots \wedge e^j),
                 \hskip 1.5cm
                 *(e^i \wedge \dots \wedge e^j),
\end{equation}
and then combine the two results into the answer to (\ref{starondecomposedalpha}).

Note: In what follows, we multiply the canonical tensors $h^{ab}, \tilde h_{ab},k_{ab},\tilde k^{ab}$
by a \emph{free parameter} $\lambda$
(and sometimes also $\omega$ and $\tilde \omega$ by $\mu$; the tensors clearly remain to be canonical)
and choose particular value of $\lambda$ (and $\mu$) at the end of computation
so that the result resembles the standard Lorentzian case \emph{as much as possible}.

\subsubsection{Hodge stars based on tensors $h^{ab}$/$\tilde h_{ab}$}
\label{HodgestarsGalCarhtildeh}

\vskip .2cm
\noindent
\textbf{Galilean case} - based on $h^{ab}$:
\vskip .3cm
\noindent
Here, w.r.t. an \emph{adapted} frame (see (\ref{galilei01a20tensorsmatrices}) and (\ref{topdegreeomegacomponents})),
\begin{equation} \label{omegaabcdviahab}
                 \omega^{a\dots b}_{\hskip .6cm c\dots d}
              := h^{a\hat a} \dots h^{b\hat b}\omega_{\hat a\dots \hat bc\dots d}
               = h^{a\hat a} \dots h^{b\hat b}\epsilon_{\hat a\dots \hat bc\dots d} \ .
\end{equation}
From $h^{00} = h^{0i}=0$ we can immediately deduce that
\begin{equation} \label{omega^0dots=0}
                 \omega^{0i\dots j}_{\hskip .7cm k\dots l} = 0,
                 \hskip 1cm \text{i.e.} \hskip 1cm
                 *(e^0 \wedge \hat s) = 0.
\end{equation}
On strictly \emph{spatial} $p$-form $\hat r$ it is more interesting.
For $h^{ij} = \lambda \delta^{ij}$ we get from (\ref{*nabaze}) on \emph{basis} spatial $p$-form
$$
  \begin{array} {rcl}
       * ( \underbrace{e^i \wedge \dots \wedge e^j}_p )
     &=& \frac{1}{(1+n-p)!}\omega^{i \dots j}_{\hskip .6cm c \dots d} \ e^c \wedge \dots \wedge e^d  \\
     &=& \frac{(1+n-p)}{(1+n-p)!}  \omega^{i \dots j}_{\hskip .6cm 0k \dots l} \ e^0 \wedge e^k \wedge \dots \wedge e^l  \\
     &=& \lambda^p e^0 \wedge \frac{1}{(n-p)!} \delta^{i\hat i}\dots \delta^{j\hat j}\omega_{\hat i \dots \hat j0k \dots l} \ e^k \wedge \dots \wedge e^l  \\
     &=& \lambda^p e^0 \wedge \frac{1}{(n-p)!} \epsilon_{i \dots j0k \dots l} \ e^k \wedge \dots \wedge e^l  \\
     &=& (-\lambda)^p e^0 \wedge \frac{1}{(n-p)!} \epsilon_{0i \dots jk \dots l} \ e^k \wedge \dots \wedge e^l  \\
     &=& (-\lambda)^p e^0 \wedge \frac{1}{(n-p)!} \epsilon_{i \dots jk \dots l} \ e^k \wedge \dots \wedge e^l  \\
     &=& (-\lambda)^p e^0 \wedge \hat * \left( e^i \wedge \dots \wedge e^j \right),
  \end{array}
$$
i.e., on \emph{complete} spatial $p$-form $\hat r$,
\begin{equation} \label{*onhatrGalhab}
                 *\hat r = (-\lambda)^p e^0 \wedge \hat * \hat r.
\end{equation}
Combining with (\ref{omega^0dots=0}), we have
\begin{equation} \label{*onalphawithlambdaGalhab}
                 *(e^0 \wedge \hat s + \hat r) = e^0 \wedge (-\lambda)^p \hat * \hat r.
\end{equation}
In order to make this formula as similar as possible to (\ref{starondecompalphaMink}),
we have to choose
\begin{equation} \label{fixlambdaandhij}
                 \lambda = -1 \ ,
                 \hskip 1cm \text{i.e.} \hskip 1cm
                 h^{ij} = - \delta^{ij}.
\end{equation}
For this fixation of the arbitrary constant, we get
\begin{equation} \label{*onalphawithGalhab}
                 *(e^0 \wedge \hat s + \hat r) = e^0 \wedge \hat * \hat r.
\end{equation}

\vskip .2cm
\noindent
\textbf{Carrollian case} - based on $\tilde h_{ab}$:
\vskip .3cm

Here
\begin{equation} \label{omegaabcdviatildehab}
                 \omega^{a\dots b}_{\hskip .6cm c\dots d}
              := \tilde \omega^{a\dots b\hat c\dots \hat d} \tilde h_{c\hat c} \dots \tilde h_{d\hat d}
               = \epsilon_{a\dots b\hat c\dots \hat d} \tilde h_{c\hat c} \dots \tilde h_{d\hat d}
               = \epsilon_{a\dots bk\dots l} \tilde h_{ck} \dots \tilde h_{dl} \ .
\end{equation}
From $\tilde h_{00} = \tilde h_{0i}=0$ (see (\ref{Carrollmatrixtildeh})) we can immediately deduce that
\begin{equation} \label{omega^idotsjdots=0}
                 \omega^{i\dots j}_{\hskip .7cm c\dots d} = 0,
                 \hskip 1cm \text{i.e.} \hskip 1cm
                 * \hat r = 0.
\end{equation}
On \emph{combined} $p$-form $e^0\wedge \hat s$ it is more interesting.
For $\tilde h_{ij} = \lambda \delta_{ij}$ we get (see (\ref{*nabaze})) on \emph{basis} combined $p$-form
$$
  \begin{array} {rcl}
       * ( e^0 \wedge \overbrace{e^i \wedge \dots \wedge e^j}^{p-1} )
     &=& \frac{1}{(1+n-p)!}\omega^{0i \dots j}_{\hskip .6cm c \dots d} \ e^c \wedge \dots \wedge e^d  \\
     &=& \frac{1}{(1+n-p)!}\omega^{0i \dots j}_{\hskip .6cm k \dots l} \ e^k \wedge \dots \wedge e^l  \\
     &=& \frac{\lambda^{1+n-p}}{(1+n-p)!}\epsilon_{0i \dots jk \dots l} \ e^k \wedge \dots \wedge e^l \\
     &=& \lambda^{1+n-p} \ \frac{1}{(n-(p-1))!}\epsilon_{i \dots jk \dots l} \ e^k \wedge \dots \wedge e^l \\
     &=& \lambda^{n-(p-1)} \ \hat *(e^i \wedge \dots \wedge e^j),
  \end{array}
$$
i.e.
\begin{equation} \label{*onE0hatsCartildehab}
                 *(e^0\wedge \hat s) = \lambda^{n-(p-1)} \ \hat * \hat s.
\end{equation}
Combining with (\ref{omega^idotsjdots=0}), we have
\begin{equation} \label{*onalphawithlambdaCarhab}
                 *(e^0 \wedge \hat s + \hat r) = \lambda^{n-(p-1)} \ \hat * \hat s.
\end{equation}
In order to make this formula as similar as possible to (\ref{starondecompalphaMink}),
we need to fulfill
\begin{equation} \label{needfulfill1}
                 \lambda^{n-(p-1)} = (-1)^{p-1}.
\end{equation}
This is not possible for any choice of $\lambda$. But we can use freedom in \emph{both} $\tilde \omega$ \emph{and} $\tilde h$
and take, from the very beginning,
\begin{equation} \label{omega0ijklpropor}
                 \tilde \omega^{0i\dots jk\dots l} = \mu (n) \epsilon_{0i \dots jk \dots l},
                 \hskip 1cm
                 \tilde h_{ij} = \lambda (n,p) \delta_{ij}.
\end{equation}
Then we need to fulfill (instead of (\ref{needfulfill1}))
\begin{equation} \label{needfulfill2}
                 \mu \lambda^{n-(p-1)} \equiv (\mu \lambda^n)\lambda^{-(p-1)} = (-1)^{p-1},
\end{equation}
with solution
\begin{equation} \label{needfulfill3}
                 \mu = (-1)^n
                 \hskip 1cm
                  \lambda = -1.
\end{equation}
Then we get from (\ref{*onalphawithlambdaCarhab}) the final expression
\begin{equation} \label{*onalphawithCartildehab}
                 *(e^0 \wedge \hat s + \hat r) = \hat * \hat \eta \hat s.
\end{equation}

\subsubsection{Hodge stars based on tensors $k_{ab}$/$\tilde k^{ab}$}
\label{HodgestarsGalCarktildek}

\vskip .2cm
\noindent
\textbf{Galilean case} - based on $k_{ab}$:
\vskip .3cm

Here
\begin{equation} \label{omegaabcdviakab1}
                 \omega^{a\dots b}_{\hskip .6cm c\dots d}
              := {\tilde \omega}^{a\dots b\hat c\dots \hat d} k_{c\hat c} \dots k_{d\hat d}
               = {\epsilon}_{a\dots b\hat c\dots \hat d} k_{c\hat c} \dots k_{d\hat d}.
\end{equation}
From $k_{00} = \lambda$ and all other possibilities vanishing (see (\ref{Galileimatrixk})) we have
\begin{equation} \label{omegaabcdviakab2}
                 \omega^{a\dots b}_{\hskip .6cm c\dots d}
               = {\epsilon}_{a\dots b0\dots 0} k_{c0} \dots k_{d0},
\end{equation}
and, consequently,
\begin{equation} \label{omegaabcdviakab3}
                 *(e^a \wedge \dots \wedge e^b) = \frac{1}{(1+n-p)!}
                 \lambda^{1+n-p}{\epsilon}_{a\dots b0\dots 0} \overbrace{e^0 \wedge \dots \wedge e^0}^{1+n-p}.
\end{equation}
So there are just two \emph{non-zero} cases (values of $p$):
\vskip .2cm
\noindent
1. $1+n-p = 0$ (i.e. $p=1+n$) and then
\begin{equation} \label{*e^0e^1e^nGal}
                 *(e^0 \wedge e^1 \wedge \dots \wedge e^n)
               = \epsilon_{01\dots n} = 1,
\end{equation}
\emph{irrespective} of the choice of $\lambda$.
\vskip .2cm
\noindent
2. $1+n-p = 1$ (i.e. $p=n$) and then
\begin{equation} \label{*e^1e^n1}
                 *(e^1 \wedge \dots \wedge e^n)
               = \lambda \epsilon_{1\dots n0} e^0 = (-1)^n\lambda e^0.
\end{equation}
Here we can obtain, by appropriate choice of $\lambda$, coincidence with formula (\ref{*onalphawithGalhab}).
Indeed, since (in $n$-dimensional \emph{Euclidean} space)
\begin{equation} \label{hat*e^1e^n}
                 \hat *(e^1 \wedge \dots \wedge e^n) = \sgn \hat g = 1,
\end{equation}
we can rewrite (\ref{*e^1e^n1}) as
\begin{equation} \label{*e^1e^n2}
                 *(e^1 \wedge \dots \wedge e^n)
               = (-1)^n\lambda e^0 \wedge \hat * (e^1 \wedge \dots \wedge e^n),
\end{equation}
and this becomes, for the choice $\lambda = (-1)^n$,
\begin{equation} \label{*e^1e^n3}
                 *(e^1 \wedge \dots \wedge e^n)
               = e^0 \wedge \hat * (e^1 \wedge \dots \wedge e^n),
\end{equation}
coinciding with what (\ref{*onalphawithGalhab}) says for this case.

If we denote the ``spatial volume element'' as
\begin{equation} \label{defhatomega}
                 \hat \omega := e^1 \wedge \dots \wedge e^n,
\end{equation}
we can summarize (\ref{*e^0e^1e^nGal}) and (\ref{*e^1e^n3}) as
\begin{equation} \label{*hatomegaande^0hatomega}
                 * \ \hat \omega = e^0 \wedge \hat * \hat \omega \equiv e^0
                 \hskip 1cm
                 *(e^0 \wedge \hat \omega) = 1,
\end{equation}
or, make complete summarizing as
\begin{align} %{rcl}
      \label{*Galkabp<n}
      *(e^0 \wedge \hat s + \hat r) =& \ 0                                         & \text{for}& &  &p<n,  \\
      \label{*Galkabp=n}
                    *\hat \omega    =& \ e^0 \hskip 1cm *(e^0 \wedge \hat s) = 0   & \text{for}& &  &p=n,  \\
      \label{*Galkabp=n+1}
      *(e^0 \wedge \hat \omega )    =& \ 1                                         & \text{for}& &  &p=1+n.
\end{align}

\vskip .2cm
\noindent
\textbf{Carrollian case} - based on $\tilde k^{ab}$:
\vskip .3cm

Here
\begin{equation} \label{omegaabcdviatildekab1}
                 \omega^{a\dots b}_{\hskip .6cm c\dots d}
              := \tilde k^{a\hat a} \dots \tilde k^{b\hat b} \omega_{\hat a\dots \hat b c\dots d}
               = \tilde k^{a\hat a} \dots \tilde k^{b\hat b} \epsilon_{\hat a\dots \hat b c\dots d}.
\end{equation}
From $\tilde k^{00} = \lambda$ and all other possibilities vanishing (see (\ref{Carrollmatrixtildek})) we have
\begin{equation} \label{omegaabcdviatildekab2}
                 \omega^{a\dots b}_{\hskip .6cm c\dots d}
               = \tilde k^{a0} \dots \tilde k^{b0} \epsilon_{0\dots 0 c\dots d},
\end{equation}
and, consequently,
\begin{equation} \label{omegaabcdviatildekab3}
                 *(e^a \wedge \dots \wedge e^b) = \frac{1}{(1+n-p)!}
                 \tilde k^{a0} \dots \tilde k^{b0} \epsilon_{0\dots 0 c\dots d} \overbrace{e^c \wedge \dots \wedge e^d}^{1+n-p}.
\end{equation}
So there are just two \emph{non-zero} cases (values of $p$):
\vskip .2cm
\noindent
1. $p = 0$, and then
\begin{equation} \label{*e^0e^1e^nCar}
                 *1
               \ = \ \frac{1}{(1+n)!} \epsilon_{c\dots d} \overbrace{e^c \wedge \dots \wedge e^d}^{1+n}
               \ = \ e^0 \wedge e^1 \wedge \dots \wedge e^n
               \ \equiv \ e^0 \wedge \hat \omega,
\end{equation}
\emph{irrespective} of the choice of $\lambda$.
\vskip .2cm
\noindent
2. $p = 1$, and then
\begin{equation} \label{*e^01}
                 *e^0 \ = \ \frac{1}{n!} \tilde k^{00} \epsilon_{0i\dots j} \overbrace{e^i \wedge \dots \wedge e^j}^n
                      \ = \ \lambda e^1 \wedge \dots \wedge e^n
                      \ \equiv \ \lambda \hat \omega.
\end{equation}
Here we can obtain, by the choice $\lambda = 1$, coincidence with formula (\ref{*onalphawithCartildehab}).
So
\begin{equation} \label{*e^02}
                 *e^0 = \hat \omega \ .
\end{equation}
We can summarize (\ref{*e^0e^1e^nCar}) and (\ref{*e^1e^n3}) as
\begin{equation} \label{*1ande^0}
                 * 1 =  e^0 \wedge \hat \omega,
                 \hskip 1cm
                 *e^0 = \hat \omega,
\end{equation}
or, make complete summarizing as
\begin{align} %{rcl}
      \label{*Cartildekabp=0}
      *1                            =& \ e^0 \wedge \hat \omega            & \text{for}& &  &p=0,  \\
      \label{*Cartildekabp=1}
                    *e^0            =& \ \hat \omega \hskip 1cm  *e^i = 0  & \text{for}& &  &p=1,            \\
      \label{*Cartildekabp=2dots}
      *(e^0 \wedge \hat s + \hat r) =& \ 0                                 & \text{for}& &  &p>1.
\end{align}

\subsection{\label{app:galileiexplformulae}Explicit formulas for $1+3$}

In the usual $1+3$ dimensional Minkowski, Galilei and Carroll spacetimes,
we can use standard Cartesian coordinate basis as an adapted frame
\begin{equation} \label{adaptedcoordframe}
                 (e^0,e^i) \ = \ (dt,dx^i)
                           \ = \ (dt,dx,dy,dz).
\end{equation}
Then the general decomposition formula from (\ref{alphadecomp}) reads
\begin{equation} \label{alphadecompapp}
                 \alpha = dt \wedge \hat s + \hat r,
\end{equation}
and may be further specified, for the five relevant degrees of differential forms,
as follows:
\begin{eqnarray} %{rcl}
      \label{Omega0}
             \Omega^0 : \ \ \  \alpha
                  &=& \ f,                                                                 \\
      \label{Omega1}
             \Omega^1 : \ \ \ \alpha
                  &=& \ fdt + \mathbf a \cdot d\mathbf r,                                  \\
      \label{Omega2}
             \Omega^2 : \ \ \ \alpha
                  &=& \ dt \wedge \mathbf a \cdot d\mathbf r + \mathbf b \cdot d\mathbf S, \\
      \label{Omega3}
             \Omega^3 : \ \ \ \alpha
                  &=& \ dt \wedge \mathbf a \cdot d\mathbf S + fdV,                        \\
      \label{Omega4}
             \Omega^4 : \ \ \ \alpha
                  &=& \ fdt \wedge dV.
\end{eqnarray}
where
\begin{eqnarray} %{rcl}
      \label{adr}
             \mathbf a \cdot d\mathbf r
                  &=& \ a_xdx + a_ydy + a_zdz,                                   \\
      \label{adS}
             \mathbf a \cdot d\mathbf S
                  &=& \ a_xdS_x + a_ydS_y + a_zdS_z,                             \\
             {}
                  &\equiv& \ a_xdy\wedge dz + a_ydz \wedge dx + a_zdx \wedge dy, \\
      \label{dV}
             dV
                  &=& \ dx \wedge dy \wedge dz.
\end{eqnarray}
So, concerning this presentation of forms alone,
there is no difference between the three spacetimes
(for the Minkowski case, see e.g. Sec.16.1 in \cite{Feckoangl}).

Due to (\ref{starMinkowski1}) - (\ref{starCarroll1})
and well-known Euclidean 3D-results,
\begin{equation} \label{E3hat*}
                 \hat * f = fdV,
                 \hskip .8cm
                 \hat * (\mathbf a \cdot d\mathbf r) = \mathbf a \cdot d\mathbf S,
                 \hskip .8cm
                 \hat * (\mathbf a \cdot d\mathbf S) = \mathbf a \cdot d\mathbf r,
                 \hskip .8cm
                 \hat * (fdV) = f
\end{equation}
(see e.g. Sec.8.5 in \cite{Feckoangl}) we get, in this language, the following explicit results for
          the action of the three Hodge star operators:
\footnote{The signs in the Galilei and Carroll cases were chosen so that the results resembled
          the standard Minkowski results as much as possible.}
\vskip .2cm
\noindent
\emph{Minkowski} Hodge star:
\begin{eqnarray} %{rcl}
      \label{*MinkOmega0}
             *f
                  &=& \ fdt \wedge dV,                                                      \\
      \label{*MinkOmega1}
             *(fdt + \mathbf a \cdot d\mathbf r)
                  &=& \ dt \wedge \mathbf a \cdot d\mathbf S + fdV,                         \\
      \label{*MinkOmega2}
             *(dt \wedge \mathbf a \cdot d\mathbf r + \mathbf b \cdot d\mathbf S)
                  &=& \  dt \wedge \mathbf b \cdot d\mathbf r - \mathbf a \cdot d\mathbf S, \\
      \label{*MinkOmega3}
             *(dt \wedge \mathbf a \cdot d\mathbf S + fdV)
                  &=& \ fdt + \mathbf a \cdot d\mathbf r,                                   \\
      \label{*MinkOmega4}
             *(fdt \wedge dV)
                  &=& \ -f.
\end{eqnarray}
\vskip .2cm
\noindent
\emph{Galilei} Hodge star:
\begin{eqnarray} %{rcl}
      \label{*GalOmega0}
             *f
                  &=& \ fdt \wedge dV                                              \\
      \label{*GalOmega1}
             *(fdt + \mathbf a \cdot d\mathbf r)
                  &=& \ dt \wedge \mathbf a \cdot d\mathbf S,                      \\
      \label{*GalOmega2}
             *(dt \wedge \mathbf a \cdot d\mathbf r + \mathbf b \cdot d\mathbf S)
                  &=& \  dt \wedge \mathbf b \cdot d\mathbf r,                     \\
      \label{*GalOmega3}
             *(dt \wedge \mathbf a \cdot d\mathbf S + fdV)
                  &=& \  fdt,                                                      \\
      \label{*GalOmega4}
             *(fdt \wedge dV)
                  &=& \ -f.
\end{eqnarray}
\vskip .2cm
\noindent
\emph{Carroll} Hodge star:
\begin{eqnarray} %{rcl}
      \label{*CarOmega0}
             *f
                  &=& \ fdt \wedge dV,                                            \\
      \label{*CarOmega1}
             *(fdt + \mathbf a \cdot d\mathbf r)
                  &=& \  fdV,
                                                                                  \\
      \label{*CarOmega2}
             *(dt \wedge \mathbf a \cdot d\mathbf r + \mathbf b \cdot d\mathbf S)
                  &=& \   - \mathbf a \cdot d\mathbf S,                           \\
      \label{*CarOmega3}
             *(dt \wedge \mathbf a \cdot d\mathbf S + fdV)
                  &=& \  \mathbf a \cdot d\mathbf r,                              \\
      \label{*CarOmega4}
             *(fdt \wedge dV)
                  &=& \ -f.
\end{eqnarray}
Notice that in expressions (\ref{*GalOmega4}) and (\ref{*CarOmega0}), Hodge stars based on tensors
$k_{ab}$/$\tilde k^{ab}$ were used
(i.e. (\ref{*Galkabp=n+1}) and (\ref{*Cartildekabp=0}) rather than
 (\ref{starGalilei1}) and (\ref{starCarroll1}), leading to zero result).
See the discussion at the end of Section \ref{galileicarrollHodgeformula}.

As for general Galilean and Carrollian (rather than Galilei and Carroll) spacetimes,
the same formulae are true if we replace interpretation of the l.h.s. of (\ref{adr}) - (\ref{dV})
with more general expressions
\begin{eqnarray} %{rcl}
      \label{adr2}
             \mathbf a \cdot d\mathbf r
                  &=& \ a_1e^1 + a_2e^2 + a_3e^3,                                 \\
      \label{adS2}
             \mathbf a \cdot d\mathbf S
                  &=& \ a_1e^2\wedge e^3 + a_2e^3 \wedge e^1 + a_3e^1 \wedge e^2, \\
      \label{dV2}
             dV
                  &=& \ e^1 \wedge e^2 \wedge e^3.
\end{eqnarray}
where $e^a \equiv (e^0,e^i)$, $i=1,2,3$, represent any \emph{adapted} coframe fields.

\section*{Acknowledgement}
The author acknowledges support from grants VEGA 1/0703/20 and VEGA 1/0719/23.


\begin{thebibliography}{99}




\bibitem{BambergSternberg} P.\ Bamberg, S.\ Sternberg: {\em A Course in Mathematics for Students of Physics},
\newline \indent
                             Cambridge University Press, Cambridge, 1990


\bibitem{Schutz} B.\ F.\ Schutz: {\em Geometrical Methods of Mathematical Physics},
\newline \indent
                             Cambridge University Press, Cambridge, 1982



\bibitem{CrampinPirani} M.\ Crampin, F.\ A.\ E.\ Pirani: {\em Applicable Differential Geometry},
\newline \indent
                             Cambridge University Press, Cambridge, 1986




\bibitem{GockSchuck} M.\ G\"ockeler, T.\ Sch\"ucker: {\em Differential Geometry, Gauge Theories and Gravity},
\newline \indent
                             Cambridge University Press, Cambridge, 1987



\bibitem{Frankel123l} T.\ Frankel: {\em The Geometry of Physics, An Introduction},
\newline \indent
                Cambridge University Press 1997 (1-st Ed), 2004 (2-nd Ed), 2012 (3-rd Ed)


\bibitem{Trautman} A.\ Trautman: {\em Differential Geometry for Physicists},
\newline \indent
                Bibliopolis, Napoli, 1984

\bibitem{Feckoangl} M.\ Fecko: {\em Differential Geometry and Lie Groups for Physicists},
\newline \indent
                Cambridge University Press, Cambridge, 2006 (paperback 2011)






\bibitem{DuvGibHorZha2014} Ch.\ Duval, G.\ W.\ Gibbons, P.\ A.\ Horvathy and P.\ M.\ Zhang:
\newline \indent
                Carroll versus Newton and Galilei:
                two dual non-Einsteinian concepts of time,
\newline \indent
 {\em Classical and Quantum Gravity}, {\bf 31}, 8 (2014);
% \newline \indent
                arXiv:1402.0657 [gr-qc]


\bibitem{BleekenYunus2016} D.\ Van den Bleeken and C.\ Yunus:
\newline \indent
                Newton-Cartan, Galileo-Maxwell and Kaluza-Klein,
\newline \indent
                {\em Classical and Quantum Gravity}, {\bf 33}, 13 (2016);
%\newline \indent
                arXiv:1512.03799v2 [math-ph]



\bibitem{Figueroa2019} J.\ Figueroa-O'Farrill, R.\ Grassie, S.\ Prohazka:
%\newline \indent
                 Geometry and BMS Lie algebras of spatially isotropic homogeneous spacetimes,
\newline \indent
                  {\em Journal of High Energy Physics} {\bf 119} (2019); arXiv:2009.01948 [hep-th]


\bibitem{Figueroa2020} J.\ Figueroa-O'Farrill:
%\newline \indent
                 On the intrinsic torsion of spacetime structures,
\newline \indent
                 Report number EMPG-20-14; arXiv:2009.01948 [hep-th] (2020)


\bibitem{Figueroa2022} J.\ Figueroa-O'Farrill:
%\newline \indent
                 Non-lorentzian spacetimes,
\newline \indent
                 {\em Differential Geometry and its Applications}, Volume 82, June 2022, 101894
\newline \indent
                 arXiv:2204.13609 [math.DG] 


\bibitem{Figueroa2023} J.\ Figueroa-O'Farrill and A.\ P\'erez and S.\ Prohazka:
\newline \indent
                 Carroll/fracton particles and their correspondence,
\newline \indent
                 {\em Journal of High Energy Physics} {\bf 207} (2023);
%\newline \indent
                 arXiv:2305.06730 [hep-th]



\bibitem{Ciambellietal2019} L.\ Ciambelli, R.\ G.\ Leigh, Ch.\ Marteau and P.\ M.\ Petropoulos:
\newline \indent
                Carroll structures, null geometry, and conformal isometries,
\newline \indent
                {\em Physical Review D} {\bf 100}, 046010 (2019);  arXiv:1905.02221 [hep-th]


\bibitem{Morand2018} K.\ Morand:
\newline \indent
                 Embedding Galilean and Carrollian geometries I.
                 Gravitational waves,
\newline \indent
                {\em Journal of Mathematical Physics} {\bf 61}, 082502 (2020)
\newline \indent
                 arXiv:1811.12681v2 [hep-th]



\bibitem{Banerjeeetal2020} K.\ Banerjee, R.\ Basu, A.\ Mehra, A.\ Mohan and A.\ Sharma:
\newline \indent
                Interacting conformal Carrollian theories: Cues from electrodynamics,
\newline \indent
                {\em Physical Review D} {\bf 103}, 105001 (2021); arXiv:2008.02829v4 [hep-th]



\bibitem{HenneaxSalReb2021} M.\ Henneaux and P.\ Salgado-Rebolledo:
\newline \indent
                 Carroll contractions of Lorentz-invariant theories,
\newline \indent
                 {\em Journal of High Energy Physics} {\bf 180} (2021); arXiv:2109.06708 [hep-th]


\bibitem{GrassiePhD} R.\ Grassie:
%\newline \indent
                Beyond Lorentzian symmetry,
\newline \indent
                PhD Thesis, University of Edinburgh (2021)
\newline \indent
                arXiv:2107.09495v1 [hep-th]



\bibitem{HansenPhD} D.\ Hansen:
%\newline \indent
                Beyond Lorentzian physics,
\newline \indent
                PhD Thesis, ETH Zurich (2021)
\newline \indent
https://www.research-collection.ethz.ch/handle/20.500.11850/488630




\bibitem{BellacLevyLeblond} M.\ Le Bellac and J.-M.\ L\'evy Leblond: Galilean Electromagnetism,
\newline \indent
                {\em Il Nuovo Cimento}, Vol.{\bf 14} B, N.2, 217 - 234 (1973)





\bibitem{Trautman1963} A.\ Trautman:
%\newline \indent
                Sur la th\'eorie newtonienne de la gravitation,
\newline \indent
                {\em C.R.Acad.Sci. Paris}, t.{\bf 257}, p.617-720 (1963)


\bibitem{Trautman1966} A.\ Trautman:
\newline \indent
                Comparison of Newtonian and Relativistic Theories of Space-Time,
\newline \indent
                pp. 413--425 in: {\em Perspectives in Geometry and Relativity},
                Essays in honor of V.Hlavat\'y, ed.
                by B. Hoffmann,
%\newline \indent
                Indiana Univ. Press, Bloomington, 1966.




\bibitem{Kunzle1972} H.\ P.\ K\"unzle:
%\newline \indent
                 Galilei and Lorentz structures on spacetime: Comparison of the corresponding
                 geometry and physics,
\newline \indent
                 {\em Ann. Inst. H. Poincar\'e. Phys. Th\'eor.} {\bf 17}, 4, 337-362. (1972)


\bibitem{LevyLeblond} J.-M.\ L\'evy Leblond: Une nouvelle limite non-relativiste du groupe de Poincar\'e,
\newline \indent
                Annales de l' I.H.P. Physique th\'eorique vol. 3, no. 1, pp. 1-12, (1965)



\bibitem{Fecko3+1} M.\ Fecko:
%\newline \indent
                 On 3+1 decompositions with respect to an observer field via dif\-feren\-tial forms,
\newline \indent
                 {\em Journal of Mathematical Physics} {\bf 38}, 4542-4560 (1997)


\bibitem{FeckoAPS} M.\ Fecko:
%\newline \indent
                 Modern geometry in not-so-high echelons of physics: Case studies,
\newline \indent
                 {\em Acta Physica Slovaca} {\bf 63}, No.5, 261 - 359 (2013)



\bibitem{FeckoGalCaroperators} M.Fecko:
%\newline \indent
                               Some useful operators on differential forms in Galilean and Carrollian spacetimes,
\newline \indent
                {\em SIGMA} {\bf 19} (2023), 024, 24 pages; \hskip .3cm arXiv:2206.11138 [math-ph]







\end{thebibliography}
\end{document}